\documentclass[preprint,showpacs,preprintnumbers,amsmath,amssymb]{revtex4}

\usepackage{graphicx}
\usepackage{dcolumn}
\usepackage{bm}

\begin{document}

\preprint{APS/123-QED}

\title{High-speed kinks in a generalized discrete $\phi^4$ model}

\author{Sergey V. Dmitriev$^1$, Avinash Khare$^2$,
Panayotis G. Kevrekidis$^3$, Avadh Saxena$^4$ and Ljup\'{c}o
Had\v{z}ievski$^5$ }

\affiliation{ $^1$Institute for Metals Superplasticity Problems
RAS, 39 Khalturina, Ufa 450001, Russia \\
$^2$Institute of Physics, Bhubaneswar, Orissa 751005,
India \\
$^3$ Department of Mathematics and Statistics,
University of Massachusetts, Amherst, MA 01003-4515, USA \\
$^4$Center for Nonlinear Studies and Theoretical Division, Los
Alamos National Laboratory, Los Alamos, New Mexico 87545, USA \\
$^5$Vinca Institute of Nuclear Science, PO Box 522,11001 Belgrade,
Serbia}

\date{\today}

\begin{abstract}
We consider a generalized discrete $\phi^4$ model and demonstrate
that it can support exact moving kink solutions in the form of
tanh with an arbitrarily large velocity. The constructed exact
moving solutions are dependent on the specific value of the
propagation velocity. We demonstrate that in this class of models,
given a specific velocity, the problem of finding the exact moving
solution is integrable. Namely, this problem originally expressed
as a three-point map can be reduced to a two-point map, from which
the exact moving solutions can be derived iteratively. It was also
found that these high-speed kinks can be stable and robust against
perturbations introduced in the initial conditions.
\end{abstract}


\maketitle

\section{Introduction} \label{Introduction}

Solitary waves play important role in a great variety of
applications because they are robust against perturbations and
they can transport various physical quantities such as mass,
energy, momentum, electrical charge, and also
information \cite{infeld,peyrard}. Many
popular continuum nonlinear equations support traveling solutions,
but for their discrete analogs the existence of traveling
solutions was not systematically studied until recently. Previously,
only a few integrable lattices were known to support moving solitons, e.g.,
the Toda lattice \cite{Toda} and the Ablowitz-Ladik lattice
\cite{AL}, the former being the discrete version of the KdV
equation and the latter of the nonlinear Schr\"odinger (NLS)
equation. While integrable lattices of other key continuum models
are also available \cite{Orfanidis,Encyclo}, these integrable
situations are still rather exceptional and typically not directly
relevant to experimental settings. On the other
hand, recently several wide classes of discrete models
supporting translationally invariant (TI) {\em static or
stationary} solutions have been discovered and investigated for
the discrete Klein-Gordon equation
\cite{PhysicaD,SpeightKleinGordon,BenderTovbis,JPA2005,CKMS_PRE2005,BOP,DKY_JPA_2006Mom,
oxt1,DKYF_PRE2006,Coulomb,DKKS2007_BOP,Roy,ArXivKDS} and the
discrete NLS equation
\cite{DNLSEx,krss,pel,DKYF_JPA_2007DNLSE,DNLSE1,NewJPA}. Static
solitary waves in such lattices possess the
translational Goldstone mode
\cite{JPA2005,Roy,ArXivKDS,DNLSEx,DKYF_JPA_2007DNLSE,NewJPA},
which means that the solitary waves moving with {\em vanishing}
velocity can be regarded as the exact solutions to these discrete
equations (although there are issues for finite but small velocities
as explained e.g., in \cite{Aigner2003,Kev2006}).
Moving solutions propagating along a lattice at {\em finite}
velocity have also been found analytically in
\cite{DKYF_JPA_2007DNLSE,NewJPA} and with the help of specially
tuned numerical approaches in
\cite{Aigner2003,Kev2006,pel,Iooss2006,PelIooss2005,Oxtoby2007,Barashenkov2005,OPB2006}.
Traveling bright solitons in the NLS lattice model with saturable
nonlinearity are also very interesting
\cite{Kev2006,Ljupco1,Ljupco2}. For this model the existence of
soliton frequencies with vanishing Peierls-Nabarro energy barrier
was demonstrated. It was shown that for the specific frequencies
the static version of the model equation coincides with that of
the integrable Ablowitz-Ladik equation \cite{Ljupco2}. This
implies in the mapping analysis a possibility to reduce the
three-point map to a two-point map. However, the studied model
equation is not integrable and while the solitary waves moving
with small velocity were numerically obtained for the specific
frequencies, the question about the existence of solitary waves
moving with finite velocities remains open.

Existence of lattices supporting the static TI solutions and the
exact solutions moving with finite velocity poses a natural
question, about whether there is a relation between them. In the
present study we answer this question in the positive, creating
a link between these two classes of solutions below.

On the other hand,
as regards the solitary wave motion, another interesting question
is the existence of upper bound for the propagation velocity.
In view of Lorentz invariance, clearly the solitary waves in continuum
nonlinear equations of the Klein-Gordon type
have limitations on their propagation velocity.
One of the questions that we examine below is whether a similar restriction
exists in the corresponding discrete models. Since the Lorentz
invariance is no longer a symmetry of the discrete nonlinear systems,
strictly speaking there need not be any restriction on the propagation
velocity. However, naively one would expect that typically
the propagation velocity
in the discrete nonlinear systems would be similar to the ones supported by
their continuous counterparts and not develop supersonic values. Recently,
three of us \cite{NewJPA} confirmed this naive expectation in a generalized
discrete NLS equation. A natural question, however, is whether this always
holds true.  Although in some case examples, this has already been addressed
\cite{Aigner2003}, here we address this question (and its negative
answer) more systematically in the
framework  of the recently studied discrete
$\phi^4$ model \cite{ArXivKDS}:
\begin{eqnarray} \label{DModel}
   \frac{d^2 \phi_n}{dt^2} = \frac{1}{h^2} (\phi_{n+1}+\phi_{n-1}-2\phi_n)
   +\lambda \phi_n \nonumber \\
   -A_1\phi_n^3
   -\frac{A_2}{2}\phi_n^2(\phi_{n+1}+\phi_{n-1}) \nonumber \\
   -\frac{A_3}{2}\phi_n(\phi_{n+1}^2+\phi_{n-1}^2)
   -A_4 \phi_n \phi_{n+1} \phi_{n-1} \nonumber \\
   -\frac{A_5}{2}\phi_{n+1}\phi_{n-1}(\phi_{n+1}+\phi_{n-1})
   -\frac{A_6}{2}(\phi_{n+1}^3+\phi_{n-1}^3),
\end{eqnarray}
with the model parameters satisfying the normalization constraint
\begin{eqnarray} \label{CC}
   \sum_{k=1}^{6}A_k=\lambda\,.
\end{eqnarray}
In Eq. (\ref{DModel}), $\phi_n(t)$ is the unknown function defined
on the lattice $x_n=hn$ with the lattice spacing $h>0$. Without
loss of generality it is sufficient to consider the cases
$\lambda= 1$ and $\lambda= -1$.

Our results will be displayed as follows. Analytical results are
collected in Sec. \ref{Analytics}. Examples of the exact moving
solutions to Eq. (\ref{DModel}) are given in Sec. \ref{MovingSN}
to Sec. \ref{ExactSine}. Then, in Sec. \ref{ApproxKink}, we derive
a continuum analog of Eq. (\ref{DModel}) and find its kink
solution. In Sec. \ref{VacSpec} spectrum of the vacuum solution of Eq.
(\ref{DModel}) is obtained. The relation between the exact moving
solutions and integrable maps is established in Sec.
\ref{RelationtoTI}. Section \ref{Numerics} is devoted to the
numerical analysis of stability and robustness of moving kinks. We
present our conclusions in Sec. \ref{Conclusions}.

\section{Analytical results} \label{Analytics}

Our approach to deriving analytical solutions will be, at least
in the beginning of this section somewhat similar in spirit
to the inverse method of \cite{flach99}. That is, we will postulate
the desired solution and will identify the model parameters for
which this solution will be a valid one, as we explain in more detail
below.

\subsection{Moving sn solution} \label{MovingSN}

It is easy to show that an exact solution to Eq. (\ref{DModel}) is
\begin{equation}\label{3.2}
   \phi_n = A{\rm sn} [\beta (hn+hx_0-vt),m]\,,
\end{equation}
provided the following relations are satisfied:
\begin{eqnarray}\label{3.3}
  A_5 {\rm ns}(2h\beta,m)=-A_3 {\rm ns}(h\beta,m)\,, \nonumber \\ \quad A_1 A^2=-2m\beta^2v^2\,,
  \quad A_6=0,
\end{eqnarray}
\begin{eqnarray}\label{3.4}
  \frac{2}{A^2h^2}=2A_4{\rm ns}(h\beta,m){\rm ns}(2h\beta,m) +A_2{\rm ns}^2(h\beta,m)
  \nonumber \\
  -A_3{\rm cs}(h\beta,m){\rm ds}(h\beta,m) \nonumber \\
  -A_5[{\rm cs}(2h\beta,m){\rm ds}(2h\beta,m) -{\rm ns}^2(2h\beta,m)]\,,
\end{eqnarray}
\begin{eqnarray}\label{3.5}
  \frac{2-\lambda h^2}{A^2h^2} =A_4{\rm ns}^2(h\beta,m)
  +A_2{\rm cs}(h\beta,m){\rm ds}(h\beta,m) \nonumber \\ -A_3{\rm ns}^2(h\beta,m)
  +(1+m)\frac{\beta^2v^2}{A^2}\,.
\end{eqnarray}

In Eq. (\ref{3.2}), $A$ is the amplitude, $v$ is the propagation
velocity, $\beta$ is a parameter related to the (inverse) width,
$x_0$ is an arbitrary position shift, $0 \le m \le 1$ is the
Jacobi elliptic function (JEF) modulus. Further, ${\rm
ns}(x,m)=1/{\rm sn}(x,m), {\rm cs}(x,m)={\rm cn}(x,m)/{\rm
sn}(x,m)$ and $ds(x,m)= {\rm dn}(x,m)/{\rm sn}(x,m)$, where ${\rm
sn}(x,m)$, ${\rm cn}(x,m)$, and ${\rm dn}(x,m)$ denote standard
JEFs.

Solution parameters $x_0$ and $m$ can be chosen arbitrarily.
Equations (\ref{3.3}) to (\ref{3.5}) establish five constraints
from which one can find the three solution parameters $A$, $v$,
and $\beta$, and two constraints on model parameters $h$,
$\lambda$, and $A_i$ ($i=1,2,...,6$).

It is possible to construct some other moving JEF solutions, for
example, moving ${\rm cn}$ and ${\rm dn}$ solutions (and hence
hyperbolic pulse solutions of the ${\rm sech}$ type) but we do not
discuss these here.

\subsection{Exact moving kink solution} \label{ExactKinks}

In the limit $m \rightarrow 1$, the above moving ${\rm sn}$
solution reduces to the moving kink solution
\begin{eqnarray} \label{3.10}
   \phi(x,t)  = \tanh \left[ {\beta \left( {hn + hx_0  - vt} \right)}
   \right],
\end{eqnarray}
and the relations (\ref{3.3}) to (\ref{3.5}) take the simpler form
\begin{equation}\label{3.11}
  \beta^2=\frac{-A_1}{2 v^2}\,, \quad 2A_3=-A_5(1+T)\,,
  \quad A_6=0,
\end{equation}
\begin{equation}\label{3.12}
  \frac{2T}{h^2}=
  A_4(1+T)+A_2+\frac{A_5}{2}(1+2T-T^2)\,,
\end{equation}
\begin{equation}\label{3.13}
  \frac{(2-h^2\lambda)T}{h^2}
  =A_4+A_2(1-T)+\frac{A_5}{2}(1+T) -A_1T\,,
\end{equation}
where
\begin{equation}\label{tanh2}
  T=\tanh^2(h\beta)\,.
\end{equation}
One can see that the solution is defined only if $A_1 \le 0$.

It is worth pointing out that the static \cite{NewJPA} and the
moving JEF as well as hyperbolic soliton solutions exist in this
model in the following seven cases: (i) only $A_2$ nonzero (ii)
only $A_4$ nonzero (iii) $A_3,A_5$ nonzero (iv) $A_2,A_4$ nonzero
(v) $A_2,A_3,A_5$ nonzero (vi) $A_3,A_4,A_5$ nonzero (vii) as
discussed above $A_2,A_3,A_4,A_5$ all nonzero. Further, while in
the static case, solutions exist only if $A_1,A_6=0$, in the
moving case solutions exist only if $A_6=0, A_1 \ne 0$. It may be
noted that these conclusions are valid for ${\rm sn},{\rm cn},
{\rm dn}$ as well as $\tanh,{\rm sech}$ solutions. Moreover, while
in the static case, the kink solution exists only if $\lambda>0$,
the moving kink solution is possible even when $\lambda<0$. In
fact, it turns out that in cases (i), (ii) and (iii) kink solution
with a large velocity $v$ is possible only if $\lambda<0$.

While Eqs. (\ref{3.11}) to (\ref{3.13}) give a general set of
restrictions on the model parameters supporting the exact moving
kink, below we extract and analyze three particular cases where
the restrictions attain a very simple and transparent form.

{\em Case I.} Only $A_1$, $A_2$, and $A_4$ are nonzero. These
parameters are subject to the constraint of Eq. (\ref{CC}) and one
can take $A_1 \le 0$ and $A_4 \le A_1-\lambda +2/h^2$ as free
parameters. From equations
 (\ref{3.11}) to (\ref{3.13}) and (\ref{CC}) we get
\begin{equation}\label{A1A4kink}
  \beta^2=\frac{-A_1}{2 v^2}, \quad
  A_4 = \frac{2}{h^2} - \frac{\lambda -A_1}{T}, \quad
  A_2=\lambda -A_1 -A_4.
\end{equation}
We note that the kink velocity $v$ can have an arbitrary value in case
$\lambda>0$.

Indeed, for chosen $\lambda$ and $h$, one can take any $v$ and
then, using the expressions of Eq. (\ref{A1A4kink}), find
subsequently the inverse kink width $\beta$ and the remaining
model parameters $A_4$ and $A_2$.

As a subcase of case I, one can take only $A_1$ and $A_2$ nonzero.
From Eq. (\ref{A1A4kink}) we get
\begin{equation} \label{A1A4kink000}
   \beta^2=\frac{-A_1}{2 v^2}, \quad 2T=h^2 \lambda+2v^2 h^2 \beta^2,
   \quad A_2=\lambda -A_1\,.
\end{equation}
From here it is clear that the kink solution with large $v$ is only
possible in this case if $\lambda<0$. Similar arguments are also
valid in cases (ii) and (iii) discussed above.

{\em Case II.} Another interesting case is when ($A_2$, $A_3$, $A_5$
nonzero):
\begin{eqnarray}\label{allAikink1}
 A_2 = \lambda - \frac{2}{h^2} - A_1 ,\quad
 A_3 = \lambda - \frac{2}{h^2} - A_1 + A_4 ,\nonumber \\
 A_5 = - \lambda + \frac{4}{h^2} + A_1 - 2A_4,\quad
 A_6 = 0,
\end{eqnarray}
with $A_1$ and $A_4$ being free parameters. It is clear that the
constraint of Eq. (\ref{CC}) is satisfied for any $A_1$ and $A_4$.
Conditions (\ref{3.11}) to (\ref{3.13}) reduce to
\begin{eqnarray}\label{allAikink2}
 \beta^2=\frac{-A_1}{2 v^2}, \quad
 \quad A_5= \frac{A_1 - \lambda}{T}.
\end{eqnarray}
Thus, we have a two-parameter set of moving kinks.
Kink solution exists for $A_1<0$ and $A_4>2/h^2$.
In this case
too the kink velocity $v$ can obtain any value.

As it was found in \cite{ArXivKDS}, for
\begin{eqnarray}\label{HamCond}
   A_1=4\gamma_1 \lambda, \quad A_2=6\gamma_2 \lambda,
   \quad A_3=(1-4\gamma_1-8\gamma_2) \lambda, \nonumber \\
   A_4=A_5=0, \quad A_6=2\gamma_2 \lambda\,,
\end{eqnarray}
with arbitrary $\gamma_1$ and $\gamma_2$, the model Eq.
(\ref{DModel}) has the Hamiltonian
\begin{eqnarray}\label{Ham}
   H=\sum_n \Big[ \frac{\dot{\phi}_n^2}{2}
   + \frac{(\phi_n -\phi_{n-1})^2}{2h^2}
   +\frac{\lambda}{4} -\frac{\lambda}{2}\phi_n^2
   +\gamma_1\phi_n^4  \nonumber \\ + \gamma_2 \phi_n\phi_{n-1}(\phi_n^2 +
   \phi_{n-1}^2)
   + \left(\frac{1}{4}-\gamma_1-2\gamma_2\right) \phi_n^2\phi_{n-1}^2
   \Big],\,\,\,\,
\end{eqnarray}
and hence energy is conserved in this model. Simple analysis of
Eqs. (\ref{3.11}) to (\ref{3.13}) suggests that among the models
supporting the exact moving kink solutions there are no
Hamiltonian models.

\subsection{Exact moving trigonometric solution \\
and moving four-periodic solution} \label{ExactSine}

{\em Exact moving trigonometric solution.} The discrete model of
Eq. (\ref{DModel}) supports an exact moving solution of the form
\begin{equation} \label{Trig}
   \phi_n = A\sin[h\beta(n+x_0)-vt]\,,
\end{equation}
even when $A_i$, $i=1,...,6$ are all nonzero provided
\begin{eqnarray}
   \lambda h^2-2 +2\cos(h\beta)+ v^2 \nonumber \\
   =A^2h^2\sin^2(h\beta)[A_3-A_4+(3A_6-A_5)\cos(h\beta)]\,,
\end{eqnarray}
\begin{eqnarray}
   A_1+A_4+A_3\cos(2h\beta)+(A_2+A_5)\cos(h\beta) \nonumber \\
   +A_6\cos(h\beta)[4\cos^3(h\beta)-3]=0\,.
\end{eqnarray}
From here, one can work out specific relations in the case of
various models. For example, consider the Hamiltonian model with
the parameters satisfying Eq. (\ref{HamCond}). It is easy to check
that the moving $\sin$e solution exists in this model provided
\begin{eqnarray}
   \lambda h^2 -2 +2\cos(h\beta)+ v^2 \nonumber \\
   =A^2 \lambda h^2 \sin^2(h\beta)
   [1-4\gamma_1-8\gamma_2+6\gamma_2 \cos(h\beta)]\,,
\end{eqnarray}
\begin{eqnarray}
   4\gamma_1+(1-4\gamma_1-8\gamma_2) \cos(2h\beta)
   +8\gamma_2 \cos^3(h\beta) =0.
\end{eqnarray}

{\em Exact moving four-site periodic solution.} For $\beta h= \pi/2$
the above moving trigonometric solution reduces to the moving
four-periodic solution of the form
\begin{equation} \label{55}
   \phi_n = A \sin\left[\frac{\pi}{2}(n+x_0)-vt\right],
\end{equation}
with
\begin{equation} \label{xx1}
   A_1+A_4=A_3\,, \quad
   A^2 = \frac{\lambda h^2-2+v^2}{A_1 h^2}\,.
\end{equation}
For the Hamiltonian model, Eq. (\ref{xx1}) reduces to
\begin{eqnarray} \label{66}
   \gamma_1+\gamma_2=\frac{1}{8}\,,
   \quad A^2 = \frac{\lambda h^2-2+ v^2}{4\gamma_1 \lambda h^2}\,.
\end{eqnarray}
Even in the Hamiltonian case the trigonometric solution can have
an arbitrarily large velocity.

\subsection{Approximate moving kink solution} \label{ApproxKink}

The discrete model of Eq. (\ref{DModel}) with {\em finite} $h$ can
support the solutions varying slowly with $n$ (the same class
of solutions can be studied near the continuum limit, using
$h$ as a small expansion parameter).
 For such solutions
one can use the expansion $\phi(h) \approx \phi + h\phi_x +
(1/2)h^2\phi_{xx}$ and substitute in Eq. (\ref{DModel})
$\phi_n=\phi(0)$, $\phi_{n \pm 1}=\phi(\pm h)$ to obtain
\begin{eqnarray} \label{Cont}
  \phi _{tt}  = \phi _{xx}  + \lambda \phi  - \lambda \phi ^3  -
  \frac{1}{2}h^2 B\phi ^2 \phi _{xx}  - h^2 D\phi \phi _x^2\,,
\end{eqnarray}
where
\begin{eqnarray} \label{BD}
 B &=& A_2  + 2A_3  + 2A_4  + 3A_5  + 3A_6 \,, \nonumber \\
 D &=& A_3  - A_4  - A_5  + 3A_6 \,,
\end{eqnarray}
and Eq. (\ref{CC}) was used. For $B=D=0$ Eq.
(\ref{Cont}) reduces to the continuum $\phi^4$ equation. This
equation also stems from Eq. (\ref{Cont}) in the limit $h
\rightarrow 0$.

Equation (\ref{Cont}) supports the following kink solution
\begin{eqnarray} \label{contkink}
   \phi(x,t)  = \tanh \left[ {\beta \left( {x + x_0  - vt} \right)}
   \right],
\end{eqnarray}
provided the following two conditions are satisfied
\begin{eqnarray} \label{contkinkparam}
   \beta ^2  = \frac{\lambda }{{2\left( {1 - v^2 } \right) + h^2
   D}},\quad \,\,\,  B + D = 0.
\end{eqnarray}

This is the quasi-continuum analog of the solution of Eq. (\ref{3.10}).
Let us analyze the case of $\lambda=1$, which corresponds to the
double-well potential. For $D=B=0$ or/and $h=0$ we have the
classical continuum $\phi^4$ kink with the propagation velocity
limited as $|v|<1$. On the other hand, for nonzero $D=-B$ and
nonzero $h$, there is no limitation on the kink propagation
velocity. In other words, fast kinks are possible in Eq.
(\ref{Cont}) only in the presence of several competing nonlinear
terms. Note that in case $B,D \ne 0$, Eq. (\ref{Cont}) is
not Lorentz invariant.

For wide kinks, i.e. when $\beta \ll h$, one can use the kink
solution of Eq. (\ref{Cont}) to write down the {\it approximate}
solution to the discrete model Eq. (\ref{DModel}) of the form of
Eq. (\ref{3.10}), whose parameters are given by Eq.
(\ref{contkinkparam}). The corresponding wide, fast kinks are
possible in Eq. (\ref{DModel}) only for finite $h$ and only for
$D=-B \neq 0$.

Substituting the Hamiltonian conditions of Eq. (\ref{HamCond}) into
Eq. (\ref{BD}) we find $B=2D$ while, according to Eq.
(\ref{contkinkparam}), the approximate moving kink solution exists
at $B=-D$, i.e., it exists only in non-Hamiltonian discrete
models.

\subsection{Spectrum of vacuum} \label{VacSpec}

The discrete model of Eq. (\ref{DModel}) supports the vacuum
solutions $\phi_n= \pm 1$. The spectrum of the small-amplitude waves
(phonons) of the form $\varepsilon_n(t) \sim \exp(ikn \pm i \omega
t)$, propagating in the vacuum, is
\begin{eqnarray}\label{VacuumSpectrum}
   \omega^2 =2\lambda + 2\left( \frac{2}{h^2} -B \right)
   \sin^2\left( \frac{k}{2} \right) \,,
\end{eqnarray}
where $k$ denotes wavenumber, $\omega$ is frequency and $B$ is as
given by Eq. (\ref{BD}).

Another vacuum solution, $\phi_n=0$, supports the phonons with the
dispersion relation
\begin{eqnarray}\label{VacuumSpectrumZero}
   \omega^2 =-\lambda + \frac{4}{h^2} \sin^2\left( \frac{k}{2} \right) \,,
\end{eqnarray}
and this vacuum is stable for $\lambda<0$.

\subsection{Relation to TI models} \label{RelationtoTI}

Note that for $v=0$ the exact moving solutions Eq. (\ref{3.2}),
Eq. (\ref{3.10}), and Eq. (\ref{Trig}) reduce to the static TI
solutions, i.e., solutions which, due to the presence of arbitrary
shift $x_0$, can be placed anywhere with respect to the lattice.
TI solutions of Eq. (\ref{DModel}) were discussed in detail in
\cite{ArXivKDS}. Particularly, it was established that any TI
static solution can be obtained iteratively from a two-point
nonlinear map. Let us demonstrate that the exact moving solutions
(including JEF solutions) can also be derived from a map.

Let us generalize the ansatz Eq. (\ref{3.2}) and look for moving
solutions to Eq. (\ref{DModel}) of the form
\begin{eqnarray}\label{F}
  \phi _n  = A F(\xi)\,,
  \quad \xi=\beta (h n + h x_0  - vt)\,,
\end{eqnarray}
with constant amplitude $A$, where $F$ is such a function that
\begin{eqnarray}\label{DDF}
  \frac{d^2 \phi_n}{dt^2} = - \beta^2 v^2
  (\alpha_0 \phi_n + \alpha_1 \phi_n^3 )\,,
\end{eqnarray}
with constant coefficients $\alpha_0$, $\alpha_1$. If this is the case,
the substitution of Eq. (\ref{F}) into Eq. (\ref{DModel}) will result
in a static problem essentially identical to the static form of
Eq. (\ref{DModel}), namely, in the problem
\begin{eqnarray}\label{Stationar}
   0=\frac{2}{{h^2 }}(\phi_{n - 1} - 2\phi_n + \phi_{n + 1})
   + 2(\lambda  + \alpha_0 \beta ^2 v^2)\phi _n \nonumber \\
   - 2(A_1 - \alpha _1 \beta ^2 v^2)\phi _n^3
   - A_2\phi_n^2 (\phi _{n - 1} + \phi_{n + 1})  \nonumber \\
   - A_3 \phi_n (\phi _{n - 1}^2 + \phi_{n + 1}^2)
   - 2A_4 \phi_{n - 1} \phi_n \phi_{n + 1} \nonumber \\
   - A_5 \phi_{n - 1} \phi_{n + 1} (\phi_{n - 1} + \phi_{n + 1})
   - A_6 (\phi _{n - 1}^3  + \phi _{n + 1}^3).
\end{eqnarray}
This effective separation of the spatial and temporal part
with each of them satisfying appropriate conditions is reminiscent
of the method of \cite{tsironis}.

The Jacobi elliptic functions and some of their complexes do
possess this property. For example, Eq. (\ref{DDF}) is valid for
\begin{eqnarray}\label{DDFcoeff}
  \alpha_0=1+m\,, \quad \alpha_1 = -\frac{2m}{A^2}\,, \quad
  &{\rm for}& \quad F={\rm sn}; \nonumber \\
  \alpha_0=1-2m\,, \quad \alpha_1 = \frac{2m}{A^2}\,, \quad
  &{\rm for}& \quad F={\rm cn}; \nonumber \\
  \alpha_0=m-2\,, \quad \alpha_1 = \frac{2}{A^2}\,, \quad
  &{\rm for}& \quad F={\rm dn}; \nonumber \\
  \alpha_0=1+m\,, \quad \alpha_1 = -\frac{2}{A^2}\,, \quad
  &{\rm for}& \quad F=\frac{1}{\rm sn}; \nonumber \\
  \alpha_0=1-2m\,, \quad \alpha_1 = \frac{2(m-1)}{A^2}\,, \quad
  &{\rm for}& \quad F=\frac{1}{\rm cn}; \nonumber \\
  \alpha_0=2(2m-1)\,, \quad \alpha_1 = -\frac{2}{A^2}\,, \quad
  &{\rm for}& \quad F=\frac{\rm sndn}{\rm
  cn}.\,\,\,\,\,\,\,\,\,\,\,
\end{eqnarray}

In the limit $m=1$ we get from the first line of Eq.
(\ref{DDFcoeff})
\begin{eqnarray}\label{DDFcoeff1}
  \alpha_0=2\,, \quad \alpha_1 = -2\,, \quad
  {\rm for} \quad F={\rm tanh}\,.
\end{eqnarray}

A solution moving along the chain continuously passes through all
configurations between the on-site and the inter-site ones. If one
finds a static solution to Eq. (\ref{Stationar}) that exists for
any location with respect to the lattice, then one finds the
corresponding moving solution to Eq. (\ref{DModel}).
Various static solutions to Eq. (\ref{Stationar}) that have the
desired property of translational invariance have been recently
constructed. Let us discuss the two simple cases specified in Sec.
\ref{ExactKinks}.

{\em Case I.}

For nonzero $A_1$, $A_2$, and $A_4 = \lambda - A_1 - A_2$, Eq.
(\ref{DModel}) reduces to
\begin{eqnarray}\label{Case1b}
  \frac{d^2 \phi _n}{dt^2} =
  \frac{1}{h^2}\left( {\phi _{n - 1}  - 2\phi _n  + \phi _{n + 1}
  } \right) + \lambda \phi _n  - A_1 \phi _n^3 \nonumber \\
  -\frac{A_2}{2}\phi_n^2(\phi_{n+1}+\phi_{n-1})
  - A_4\phi _{n - 1} \phi _n \phi _{n + 1},
\end{eqnarray}
while from Eq. (\ref{Stationar}) we get
\begin{eqnarray}\label{Case1c}
  \frac{1}{h^2}(\phi _{n - 1}  - 2\phi_n + \phi_{n + 1})
  + (\lambda  + \alpha_0 \beta ^2 v^2)\phi_n \nonumber \\
  - (A_1  -\alpha_1 \beta ^2 v^2 )\phi_n^3 \nonumber \\
  -\frac{A_2}{2}\phi_n^2(\phi_{n+1}+\phi_{n-1})
  - A_4\phi_{n - 1} \phi_n \phi_{n + 1} = 0\,.
\end{eqnarray}
Setting
\begin{eqnarray}\label{Case1d}
  A_1 = \alpha_1 \beta ^2 v^2\,, \quad  \quad
  \alpha_0 = -\alpha_1\,,
\end{eqnarray}
we reduce Eq. (\ref{Case1c}) to the integrable static equation
\cite{DKKS2007_BOP}
\begin{eqnarray}\label{Case1a}
  \frac{1}{h^2}(\phi_{n - 1} - 2\phi_n  + \phi_{n + 1})
  + P\phi_n - \frac{A_2}{2}\phi_n^2(\phi_{n+1}+\phi_{n-1}) \nonumber \\
  - (P-A_2)\phi_{n - 1} \phi_n \phi _{n + 1} = 0,\,\,\,\,
\end{eqnarray}
with
\begin{eqnarray}\label{Case1e}
  P=\lambda -A_1\,.
\end{eqnarray}
Any {\em static} TI solution to Eq. (\ref{Case1a}) generates the
corresponding {\em moving} solution to Eq. (\ref{Case1b}), provided
that Eq. (\ref{DDF}) is satisfied.

All static solutions of Eq. (\ref{Case1a}) can be found from its
first integral \cite{ArXivKDS},
\begin{eqnarray}\label{qa}
   \phi_n^2&+&\phi_{n+1}^2-\frac{Y\phi_n^2\phi_{n+1}^2 } {2-Ph^2}
   -2Z\phi_n \phi_{n+1}-\frac{CY}{2-Ph^2}=0\,, \nonumber \\
   Z &=& \frac{{( 2 - Ph^2 )^2  - Ch^4  (P - A_2) ^2 }}
   {{2\left( {2 - Ph^2 } \right) + C h^4 A_2 (P - A_2) }},  \nonumber \\
   Y &=& h^2 ( P - A_2 + A_2 Z),\,\,\,\,\,\,\,
\end{eqnarray}
where $C$ is the integration constant. The first integral can be
viewed as a nonlinear map from which a particular solution can be
found iteratively for an admissible initial value $\phi_0$ and for
a chosen value of $C$.

Thus, Eq. (\ref{qa}) is a general solution to Eq. (\ref{Case1a}),
while to make it also a solution of Eq. (\ref{Case1c}) one needs
to satisfy Eq. (\ref{Case1d}). This can be achieved by calculating
$d^2\phi_n/d\xi^2$ from the map Eq. (\ref{qa}) and equalizing it
to $-\alpha_0 \phi_{n+1} - \alpha_1 \phi_{n+1}^3$ according to Eq.
(\ref{DDF}). To calculate $d^2\phi_n/d\xi^2$ we consider a static
solution centered at the lattice point with the number $n$ so that
$\phi_n=0$. From the map Eq. (\ref{qa}) one can find
$\phi_{n+1}(0)=\sqrt{CY/(2-Ph^2)}$. We then consider the initial
values $\phi_n=-\varepsilon$ and $\phi_n=\varepsilon$ and
similarly, from the map Eq. (\ref{qa}), find the corresponding
values of $\phi_{n+1}(-\varepsilon)$ and
$\phi_{n+1}(\varepsilon)$. Finally, we calculate the second
derivative at the point $n+1$ as $d^2\phi_n/d\xi^2= \mathop {\lim
}\limits_{\varepsilon ^2 \to 0} \left[ {\phi _{n + 1} \left( { -
\varepsilon } \right) - 2\phi _{n + 1} \left( 0 \right) + \phi _{n
+ 1} \left( \varepsilon \right)} \right]/\varepsilon ^2 =
-\alpha_0 \phi_{n+1} - \alpha_1 \phi_{n+1}^3$, where the values of
the coefficients are
\begin{eqnarray}\label{al0al1}
  \alpha _0  = \frac{{CY^2  - \left( {2 - Ph^2 } \right)^2 \left(
  {Z^2  - 1} \right)}}{{\left( {2 - Ph^2 }
  \right)CY}}, \quad \alpha _1  =  - \frac{2}{C}.
\end{eqnarray}

Let us summarize our findings. From Eq. (\ref{al0al1}) and the
second expression of Eq. (\ref{Case1d}), taking into account Eq.
(\ref{Case1e}), we find the relation between the integration
constant $C$ and model parameters $A_1$, $A_2$, $\lambda$, and $h$
when a moving solution is possible:
\begin{eqnarray}\label{Cparam}
   \left( {1 - C} \right)P\left( {2 - Ph^2 } \right)^2   \nonumber \\
  \times\Big[ h^4 (C-1)P^3  + Ch^4 A_2
  \left( {5A_2 P - 4P^2  - 2A_2^2} \right) \nonumber \\
  + 4\left( {2A_2 - 3P  - A_2 Ph^2  + 2P^2 h^2 } \right)\Big] = 0\,.
\end{eqnarray}

The solution profile is found for this $C$ from Eq. (\ref{qa})
iteratively for an admissible initial value $\phi_0$, and the
propagation velocity $v$ is found from the first expression of Eq.
(\ref{Case1d}) and the second expression of Eq. (\ref{al0al1}).
Following this way, {\em any} moving solution to Eq.
(\ref{Case1b}) can be constructed (possibly, except for some very
special solutions that may arise from factorized equations
\cite{ArXivKDS}). Note that in \cite{DKKS2007_BOP} we could
express many but not all the solutions of Eq. (\ref{Case1a}) in
terms of JEF. The approach developed in this section allows one to
obtain iteratively even those moving solutions whose corresponding
static problems were not solved in terms of JEF.

The simplest case is $C=1$ when Eq. (\ref{Cparam}) is satisfied
and this case corresponds to the kink solution. From Eq.
(\ref{al0al1}) we find
\begin{eqnarray}\label{Kinka0a1}
  \alpha_0=2, \quad \alpha_1=-2\,,
\end{eqnarray}
which coincides with Eq. (\ref{DDFcoeff1}). The map Eq. (\ref{qa})
in this case reduces to
\begin{eqnarray}\label{KinkMap1}
  \pm \sqrt {\frac{(2/h^2) - A_4}{\lambda - A_1}} (\phi _n  - \phi _{n + 1})
  - \phi _n \phi _{n + 1}  + 1 = 0\,.
\end{eqnarray}
One can check that Eq. (\ref{KinkMap1}) supports the static
solution $\phi_n = \tanh [h\beta (n+x_0)]$ with
\begin{eqnarray}\label{Case1j}
  {\rm{tanh}}^2(\beta h) = \frac{\lambda - A_1}{2/h^2 - A_4}\,,
\end{eqnarray}
and this coincides with the second expression of Eq.
(\ref{A1A4kink}). The static solution Eq. (\ref{Case1j}), that can
also be found iteratively from Eq. (\ref{KinkMap1}), gives the
profile of the {\em moving} kink that satisfies Eq.
(\ref{Case1b}). The kink velocity $v$ is found from Eq. (\ref{Case1d})
and Eq. (\ref{Kinka0a1}) and this agrees with the first expression
of Eq. (\ref{A1A4kink}).

{\em Case II.}

The model parameters are related by Eq. (\ref{allAikink1}) with $A_1$
and $A_4$ being free parameters. For $A_1 = - 2\beta ^2 v^2$, Eq.
(\ref{Stationar}) reduces to
\begin{eqnarray}\label{3.11xz}
 \frac{2}{{h^2 }}\left( {\phi _{n - 1}  - 2\phi _n  + \phi _{n + 1} } \right)
 + 2\left( {\lambda  - A_1 } \right)\phi _n  \nonumber \\
  - A_2 \phi _n^2 \left( {\phi _{n - 1}
  + \phi _{n + 1} } \right) - A_3 \phi _n \left( {\phi _{n - 1}^2
  + \phi _{n + 1}^2 } \right)  \nonumber \\
  - 2A_4 \phi _{n - 1} \phi _n \phi _{n + 1}
  - A_5 \phi _{n - 1} \phi _{n + 1} ( {\phi _{n - 1}  + \phi _{n + 1} } ) =
  0,\,\,\,\,
\end{eqnarray}
which is a particular form of the case (vii) static equation
studied in \cite{ArXivKDS}. The first integral of the static model
(vii) for the general case is not known, but it is known for any
particular JEF solution \cite{ArXivKDS}. Let us further simplify
the problem considering only the kink solution of Eq.
(\ref{3.11xz}), for which one has
\begin{eqnarray}\label{3.11x}
  \tanh ^2 (h\beta) = \frac{A_1 - \lambda}{A_5}\,.
\end{eqnarray}
This relation coincides with the second expression of Eq.
(\ref{allAikink2}). From the known static $\rm tanh$ solution one
can deduce the following two-point map that generates this
solution for any initial value $|\phi_0|<1$:
\begin{eqnarray}\label{3.11y}
  \phi_{n+1} = \frac{\phi_n\pm \sqrt{(A_1 - \lambda)/A_5}}
  {1\pm \phi_n \sqrt{(A_1 - \lambda)/A_5}}\,,
\end{eqnarray}
where one can interchange $\phi_n$ and $\phi_{n+1}$ and take
either the upper or the lower sign.

The static kink solution to Eq. (\ref{3.11xz}) with $\beta$
satisfying Eq. (\ref{3.11x}), that can also be found iteratively
from Eq. (\ref{3.11y}), gives the profile of the {\em moving} kink
that satisfies Eq. (\ref{DModel}) with the parameters related by
Eq. (\ref{allAikink1}). The kink velocity $v$ is found from Eq.
(\ref{Case1d}) and Eq. (\ref{Kinka0a1}).

\begin{figure}
\includegraphics{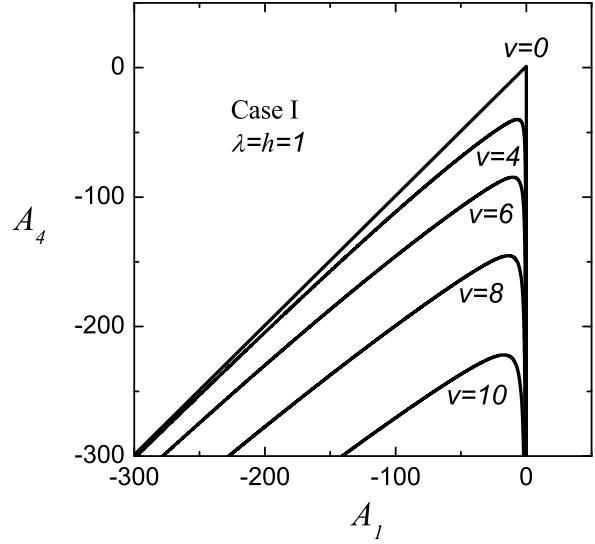}
\caption{Isolines of equal kink velocity on the plane of model
parameters $A_1$, $A_4$ for the Case I at $\lambda= h= 1$,
$A_2=\lambda-A_1-A_4$, $A_3=A_5=A_6=0$. } \label{Figure1}
\end{figure}

\begin{figure}
\includegraphics{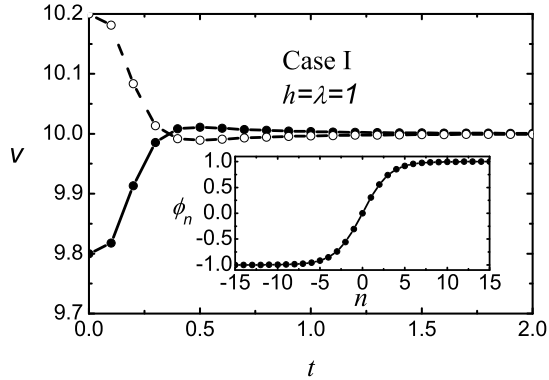}
\caption{Kink velocity as a function of time for the Case I.
The exact kink velocity for chosen model parameters is  $v=10$ but for
setting the initial conditions we put in Eq. (\ref{3.10}) $v=9.8$
(dots) and $v=10.2$ (open circles). In both cases, regardless of the
sign of perturbation, the kink velocity rapidly approaches the exact
value. The inset shows the moving kink profile. Model parameters:
$\lambda= h= 1$, $A_1=-20$, $A_4 \approx -222.1$,
$A_2=\lambda-A_1-A_4$, $A_3=A_5=A_6=0$.} \label{Figure2}
\end{figure}

\section{Numerics} \label{Numerics}

Let us analyze the kink solutions for the cases I and II described
in Sec. \ref{ExactKinks}. Only the case of $\lambda=1$ will be
analyzed.

In our simulations we solve the set of equations of motion, Eq. (\ref{DModel}), numerically with a sufficiently small time step $\tau$ using the
Stormer integration scheme of order $O(\tau^6)$. Initial
conditions are set by utilizing Eq. (\ref{3.10}) with various
parameters. Anti-periodic or fixed boundary conditions are
employed.

\subsection{Exact moving kinks in Case I} \label{KinksI}

In this case, as it was already mentioned, the exact moving kink
solution exists for the free model parameters satisfying $A_1 \le
0$ and $A_4 \le A_1-\lambda +2/h^2$ (see Fig. \ref{Figure1} where
for $\lambda=h=1$ we show the isolines of equal kink velocity on
the plane of model parameters $A_1$, $A_4$). On the line $A_1=0$,
according to the first expression of Eq. (\ref{A1A4kink}), the
kink velocity $v$ vanishes. On the line $A_4 = A_1-\lambda +2/h^2$
the kink velocity also vanishes. This is so because, as it can be seen
from the second expression of Eq. (\ref{A1A4kink}), on this line
we have $T=1$, i.e. $\beta \rightarrow \infty$ and kink width
vanishes. On the line $A_4 = A_1-\lambda +2/h^2$ we also have
$B=2/h^2$, which corresponds to vanishing of the width of the
phonon spectrum given by Eq. (\ref{VacuumSpectrum}).

The vacuum $\phi_n=\pm 1$ is stable, i.e. the spectrum Eq.
(\ref{VacuumSpectrum}) does not have imaginary frequencies, if
$A_4<A_1+2/h^2$, i.e., it is stable in the whole region where the
exact moving kink solution is defined.

As it was already mentioned, the kink propagation velocity is
unlimited and from Fig. \ref{Figure1} one can see that $|A_4|$
increases for higher kink velocities while $A_1$ can have any
negative value. For example, for $v=10$ the moving kink exists for
$A_1<0$ and $A_4<-221.9$.

Let us now turn to the discussion of the stability of moving kinks.

{\em Case of $\lambda=h=1$ and $v=10$.}

Moving kink solutions are stable not in the whole range of
parameters of their existence but for each studied value of
propagation velocity we were able to find a range of parameters
where the moving kink is stable in the sense that even a perturbed
kink solution (with a reasonably large perturbation amplitude) in
course of time tends to the exact solution (i.e., it is
effectively ``attractive''). The evolution of the kink velocity in
such self-regulated kink dynamics is shown in Fig. \ref{Figure2}
for a velocity as large as $v=10$ for $A_1=-20$ at $\lambda=h=1$.
For this choice, we have from Eq. (\ref{A1A4kink}) $\beta \approx
0.316$, $A_4 \approx -222.1$, $A_2 \approx 243.1$. The moving kink
profile is shown in the inset of Fig. \ref{Figure2}. The
perturbation was introduced into the initial conditions by setting
in the exact solution of Eq. (\ref{3.10}), a ``wrong'' propagation
velocity, smaller or higher than the exact value $v=10$. It can be
seen that in course of time the propagation velocity approaches
the exact value regardless of the sign of perturbation. The increase
of kink velocity launched with $v=9.8$ may look counterintuitive
but one should keep in mind that moving kinks are the solutions to
a non-Hamiltonian (open) system with the possibility to have energy
exchange with the surroundings with gain or loss, depending on the
trajectories of particles (see, e.g., \cite{JPA2005}).

We found that the absolutely stable, self-regulated motion of
kink, similar to that shown in Fig. \ref{Figure2}, for $v=10$ and
$\lambda=h=1$ takes place within the range of $-164< A_1< -1.1$.
The inverse kink width at the lower edge of the stability window
is $\beta \approx 0.906$, which corresponds to a rather sharp kink,
while for $A_1=-1.1$ (the upper edge of the stability window) one
has $\beta \approx 0.0742$ and the kink is much wider than the lattice
spacing $h$. For comparison, the kink shown in the inset of Fig.
\ref{Figure2} has $\beta \approx 0.316$. For $A_1< -164$ the
moving kink solution becomes unstable and displacements of
particles behind the kink grow rapidly with time resulting in the
stopping of numerical run due to floating point overflow. For $-1.1<
A_1 <0$ kink dynamics is as described in Sec. 6 of \cite{JPA2005}
for the non-Hamiltonian case. In this region of parameter $A_1$, after
a transition period, the kink starts to excite in the vacuum
in its wake a wave with constant amplitude. In this regime, the
kink attains a constant velocity whose value is, generally
speaking, different from that prescribed by Eq. (\ref{A1A4kink}).
Kink dynamics in the two unstable regimes described above will be
illustrated below for the kinks moving with $v<1$.


{\em Case of $\lambda=h=1$ and $v<1$.}

Similar results were obtained for the kinks moving with small
velocities ($v<1$). For smaller velocities the range of $A_1$ with
stable, self-regulated motion becomes narrower. For instance, a kink
with $v=0.8$ is stable (in the above mentioned sense) for $-2.4
<A_1 <-0.10$, while the one with $v=0.5$ for $-0.19 <A_1 <-0.06$.

\begin{figure}
\includegraphics{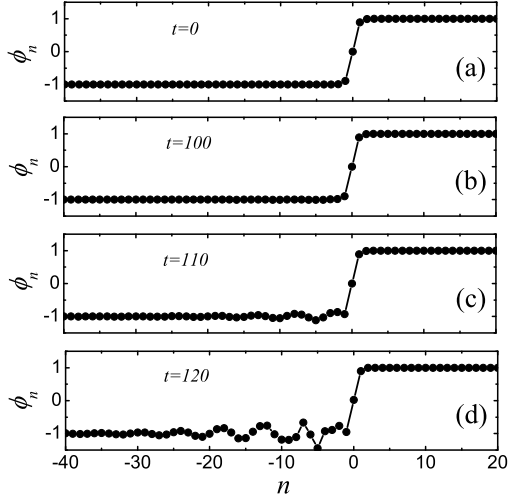} \caption{Change of the moving kink
profile demonstrating the instability of the exact solution for
the following set of model parameters: $\lambda= h= 1$,
$A_1=-2.6$, $A_4\approx -2.538$, $A_2=\lambda-A_1-A_4\approx
6.138$, $A_3=A_5=A_6=0$. Exact kink parameters are $v=0.8$,
$\beta\approx 1.425$. This solution is unstable and, due to the
presence of rounding error perturbation, the wave behind the kink
is excited. The amplitude of the wave rapidly growths with time
and it becomes noticeable in the scale of the figure at $t> 100$.
At $t\approx 130$ the numerical run stops due to the floating
point overflow.} \label{Figure3}
\end{figure}

In Fig. \ref{Figure3} we show the time variation of the moving
kink profile to demonstrate the instability of the exact kink
solution moving with $v=0.8$ at $A_1=-2.6$. This solution is
unstable and, due to the presence of rounding error perturbation,
the wave behind the kink is excited. The amplitude of the wave
rapidly grows with time and it becomes noticeable in the scale
of the figure at $t> 100$. At $t\approx 130$ the numerical run
stops due to the floating point overflow. Kink velocity is nearly
equal to $v=0.8$ practically until the collapse of the wave behind
it.

\begin{figure}
\includegraphics{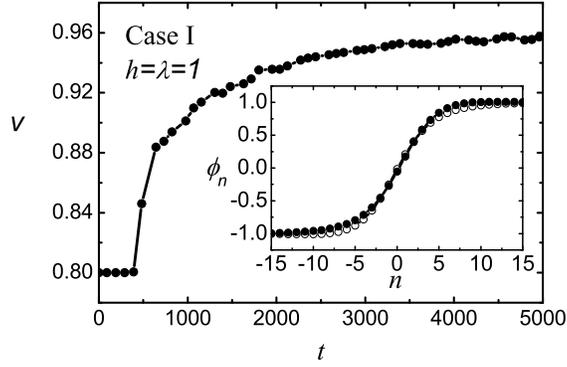}
\caption{Kink velocity as a function of time demonstrating the
instability of the exact moving kink solution for the following
set of model parameters: $\lambda= h= 1$, $A_1=-0.07$, $A_4\approx
-18.28$, $A_2=\lambda-A_1-A_4\approx 19.35$, $A_3=A_5=A_6=0$.
Exact kink parameters are $v=0.8$, $\beta\approx 0.234$. This
solution is unstable and, due to the presence of rounding error
perturbation, at $t\approx 400$ it starts to transform to an
oscillating moving kink (with the period $T\approx 5.01$). The
transformation is essentially complete by $t\approx 5000$. Inset
shows the profiles of the oscillating moving kink in the two
configurations with the most deviation from the average in time
configuration.} \label{Figure4}
\end{figure}

In Fig. \ref{Figure4} we show the kink velocity as a function of
time to illustrate the instability of the exact moving kink
solution at $v=0.8$ and $A_1=-0.07$. This solution is unstable
and, due to the presence of rounding error perturbation, at
$t\approx 400$ it starts to transform into an oscillating kink (with
the period $T\approx 5.01$) moving with different velocity and
exciting a constant-amplitude wave behind it. The transformation
is essentially complete by $t\approx 5000$ and the kink velocity
becomes $v\approx 0.96$. The inset shows the profiles of
an oscillating moving kink in the two configurations with the most
deviation from the average in time configuration.

{\em Case of $\lambda=1$, $h=0.1$, and $v=10$.}

Stable, self-regulated motion of high-speed kinks was also
observed for as small lattice spacing as $h=0.1$. This may appear
surprising at first sight because for small $h$ one would
expect the discrete model to be close to the continuum $\phi^4$
model where propagation with the velocity faster than $v=1$ is
impossible. But looking at Eq. (\ref{Cont}), one can notice that
the two last non-Lorentz-invariant terms have the coefficients
$h^2B$ and $h^2D$, and they are not small even for small $h$ if
$B$ and $D$ are large. The high-speed kink in the considered case
indeed exists for $B\approx -19833.1$ and $D\approx 20034.1$.
These values correspond to the following parameters considered in
this numerical run: $A_1=-200.0$, $A_4\approx -20034.1$,
$A_2=\lambda-A_1-A_4\approx 20235.1$, $A_3=A_5=A_6=0$, $\beta=1$.

\begin{figure}
\includegraphics{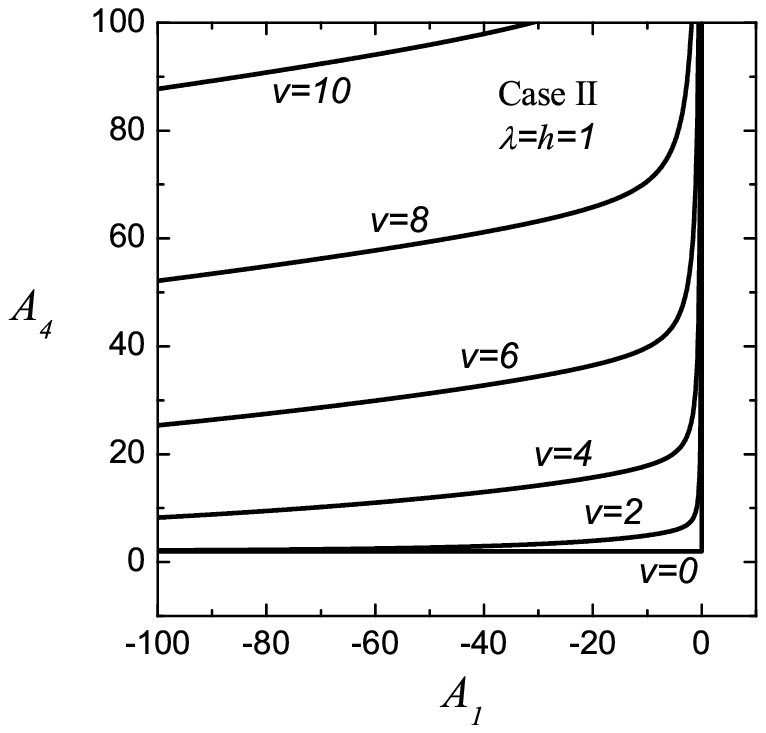}
\caption{Isolines of equal kink velocity on the plane of model
parameters $A_1$, $A_4$ at $\lambda= h= 1$ for the case II when
model parameters satisfy Eq. (\ref{allAikink1}). } \label{Figure5}
\end{figure}

\subsection{Exact moving kinks in Case II} \label{KinksII}

In the case II, the  model parameters are related by Eq.
(\ref{allAikink1}) and we find $B=-2A_4+6/h^2$. As it was already
mentioned, the exact moving kink solution exists for $A_1<0$ and
$A_4>2/h^2$. Isolines of equal kink velocity on the plane of the
free model parameters $A_1$, $A_4$ can be seen in Fig.
\ref{Figure5} for $\lambda=h=1$. In the Case II, similarly to the
Case I, on the borders of the existence of the kink solution the
kink velocity vanishes. Moreover, on the border $A_4=2/h^2$, both
the width of the phonon spectrum given by Eq.
(\ref{VacuumSpectrum}) and the kink width vanish, and this is also
similar to what we saw in Case I.

As was mentioned above, the exact moving kink solution in the Case II
can also have arbitrary velocity. Kink dynamics in the Case II was
observed to be qualitatively similar to that of the Case I. For
the kink with speed as large as $v=10$ for $\lambda=h=1$ we found
that the absolutely stable self-regulated motion is observed
within the range of $-82 < A_1 <-1.0$. The corresponding inverse
kink width varies in the range $0.640 < \beta < 0.0707$. For $A_1<
-82$ the moving kink solution becomes unstable and displacements
of particles in the wake of the kink grow rapidly with time resulting in
the termination of numerical run due to floating point overflow. For
$-1.0< A_1 <0$, after a transition period, the kink starts to excite
in the vacuum behind itself a wave with constant amplitude. In
this regime, the kink propagates with a constant velocity whose value
is, generally speaking, different from that prescribed by Eq.
(\ref{allAikink2}).

\subsection{Exact moving four-periodic solution} \label{FourP}

We have studied the dynamics of the exact moving four-site periodic
solution Eq. (\ref{55}) in the Hamiltonian model, i.e., with the
parameters satisfying Eq. (\ref{66}). Periodic boundary conditions
were employed for a chain of 40 particles. A small perturbation
was introduced in the amplitude of particles at $t=0$ and we
always observed a growth of deviation of the perturbed solution
away from the exact one. We varied the free model parameter
$\gamma_1$ and the solution parameter $v$ in wide ranges at
$\lambda=h=1$ but could not find a stable regime.

\section{Conclusions} \label{Conclusions}

For the discrete model of Eq. (\ref{DModel}), in Sec.
\ref{MovingSN} we obtained exact moving solutions in the form of
the sn Jacobi elliptic function. Solutions in the form of cn and
dn Jacobi elliptic functions can also be constructed as well as
the solutions having the form of 1/sn, 1/cn, and sndn/cn in
analogy with \cite{DKKS2007_BOP}.  In Sec. \ref{ExactKinks}, from
the sn solution, in the limit of $m \rightarrow 1$, we extracted
the exact moving kink solution in the form of tanh, i.e., the
corresponding hyperbolic function solution.


Setting $v=0$ in the exact moving solutions obtained in this work
one obtains the TI static solutions reported in \cite{ArXivKDS}.
In this sense, the results reported here generalize our previous
results. We thus reveal the hidden connection between the static
TI solutions and the exact moving solutions. Such solutions can be
derived from a three-point map reducible to a two-point map (see Sec.
\ref{RelationtoTI}) i.e., from an integrable map \cite{Quispel}.

We have demonstrated that the exact moving solutions to lattice
and continuous equations with competing nonlinear terms can have
any large propagation velocity. Most of the high-speed solutions
reported in the present study are solutions to the non-Hamiltonian
variant of the considered
$\phi^4$ models. However, the trigonometric solution described in Sec.
\ref{ExactSine} exists also in the Hamiltonian lattice and it also can
have an arbitrary speed.


While the problem of identifying traveling solutions in the
one-dimensional context by now has a considerable literature
associated with it, as evidenced above, identifying such solutions
in higher dimensional problems is to a large extent an open
question. While initial studies have demonstrated the possibility
in some of these systems for traveling in both on- and off-lattice
directions \cite{susanto}, analytical results along the lines discussed here
are essentially absent in that problem and could certainly assist
in clarifying the potential of coherent structures for unhindered
propagation in these higher dimensional settings.

\section*{Acknowledgements}
SVD gratefully acknowledges the financial support provided by the
Russian Foundation for Basic Research, grant 07-08-12152.
PGK gratefully acknowledges support from NSF-DMS, NSF-CAREER
and the Alexander-von-Humboldt Foundation.  This work was supported
in part by the U.S. Department of Energy.

\end{document}